\documentstyle[11pt, aaspp4,flushrt]{article}
\begin{document}
\title{Recovery of the X-Ray Transient QX~Nor (=X1608-52)
 in Outburst and Quiescence}
\author{Stefanie Wachter\altaffilmark{1}}  
\affil{Department of Astronomy, University of Washington, Box 351580,
 Seattle, WA 98195-1580; wachter@astro.washington.edu}
\altaffiltext{1}{Visiting Astronomer, Cerro Tololo Interamerican Observatory,
National Optical Astronomy Observatories, operated by AURA, Inc. under 
cooperative agreement with the NSF.}
\authoremail{wachter@astro.washington.edu}

\begin{abstract}

We present optical and near-IR observations of QX~Nor, the counterpart to the
recurrent soft X-ray transient X1608-52, after its reappearance following
the X-ray outburst in February 1996. The object has been seen only once 
before, during an X-ray outburst in 1977.
Data from 3--5 months after the outburst 
show the counterpart at a mean magnitude of  
$R=20.2$ and variable on timescales of days. 
A comparison with identical observations in 1995 implies 
that the object has brightened by at least 1.8~mag in $R$ following the 
X-ray outburst.  
We also detected QX~Nor in the IR in {\it both} 
quiescence and outburst. A faint source is visible in the $J$ but not the $R$ 
band in May 1995. These first observations in the quiescent state yield 
magnitudes and colors consistent with 
optical emission from a low mass companion in the binary system, as is true in 
other soft X-ray transients. 
 
\end{abstract}


\section{Introduction}

X1608-52 is a soft X-ray transient (SXT) that exhibits large X-ray
outbursts lasting tens of days on irregular timescales ($\sim$100~d to years, 
Lochner \& Roussel-Dupre 1994). X1608-52 is an unusual source since
the detection of type I X-ray bursts places it into the rare group
of SXT with neutron star primaries; most SXTs are black hole candidates. 
The only other members of this group 
with optical counterparts are Cen~X-4 and Aql~X-1. (For a review on SXTs see 
van Paradijs \& McClintock 1995). Furthermore, it is one of the few SXTs
with a bright quiescent X-ray flux, allowing for an in depth study of 
its X-ray properties
in both quiescence and outburst (Fujimoto \& Gottwald 1989; 
Mitsuda et al. 1989; Yoshida et al. 1993). 
It has been classified
as an Atoll-source by Hasinger \& van der Klis (1989) and also shows 
some characteristics that are usually associated with  
black hole candidates (Tanaka \& Lewin 1995). Most recently, an 
800~Hz QPO has also been detected 
(Berger et al. 1996).
 
The X-ray outbursts in SXTs are thought to be due to enhanced
accretion caused by an instability in either the secondary star 
(Hameury, King, \& Lasota 1986) or the 
accretion disk (Lin \& Taam 1984; Mineshige \& Wheeler 1989). 
The optical counterpart of X1608-52, QX~Nor, was discovered by Grindlay \& 
Liller (1978) after an outburst in 1977. During quiescence, the  
counterpart fades to as yet undetected faint optical magnitudes. 
We present observations of QX~Nor after its reappearance following 
the X-ray outburst detected with XTE in February 1996 
(Marshall et al. 1996).

\section{Observations}

\subsection{Optical Photometry}

CCD $R$ and $I$ band photometry of QX~Nor was performed with the CTIO  
0.9~m telescope on 1995 May 28 UT and on several occasions in 1996 May, June, 
and July.
Overscan and bias corrections were made for each CCD image with the 
task {\it quadproc} at CTIO to deal with the 4 amplifier readout.  
The data were flat-fielded in the standard 
manner with IRAF.
 
\begin{figure}[h]
\plotone{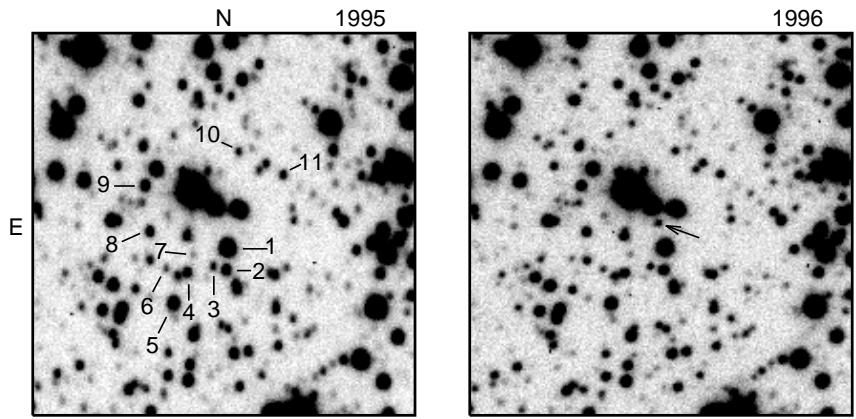}
\caption[]{{\it Left:} 1995 600~s $R$ band exposure of the
field of X1608-52 taken with the CTIO 0.9~m telescope.
The field size is 90\arcsec$\times$90\arcsec. All comparison
stars are marked.
{\it Right:} 1996 600~s $R$ band exposure (MJD 50,276.1) of the same field. The
arrow
marks QX~Nor. \label{f-fc}}
\end{figure}

Since there is only a single previously published finding chart of QX~Nor 
(Grindlay \& Liller 1978), 
obtained with photographic plates
and of modest quality,   
a CCD image of the field from both 1995 and 1996 is displayed in 
Figure~\ref{f-fc}. All comparison stars used in the analysis 
are marked.

Due to the crowded field, all
photometry was performed by point spread function fitting with DAOPHOT~II
(Stetson 1993). 
The instrumental magnitudes were transformed to the standard system through
observations of several Landolt standard star fields 
(Landolt 1992). 
The magnitudes of QX~Nor and several local
comparison stars are listed in Table~\ref{t1}. The systematic error (from the 
transformation to the standard system) in these optical	
magnitudes is $\pm 0.06$~mag. The intrinsic 1$\sigma$ error of the relative
photometry was derived from the rms scatter in the lightcurve of the comparison
stars. The errors vary between $\pm 0.01$ and $\pm 0.05$~mag depending on the 
brightness of the stars (see Figure~\ref{f-lc}). 

\begin{figure}[ht]
\plotfiddle{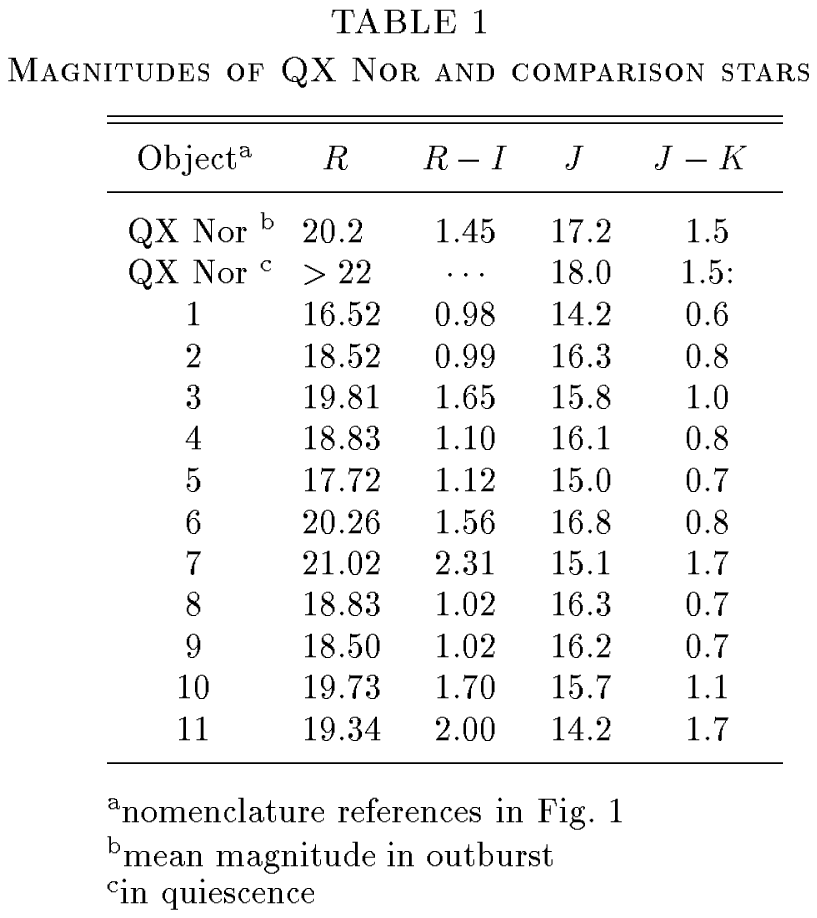}{10.0cm}{0}{100}{100}{-324}{-252}
\end{figure}

 \begin{figure}[th]
\plotfiddle{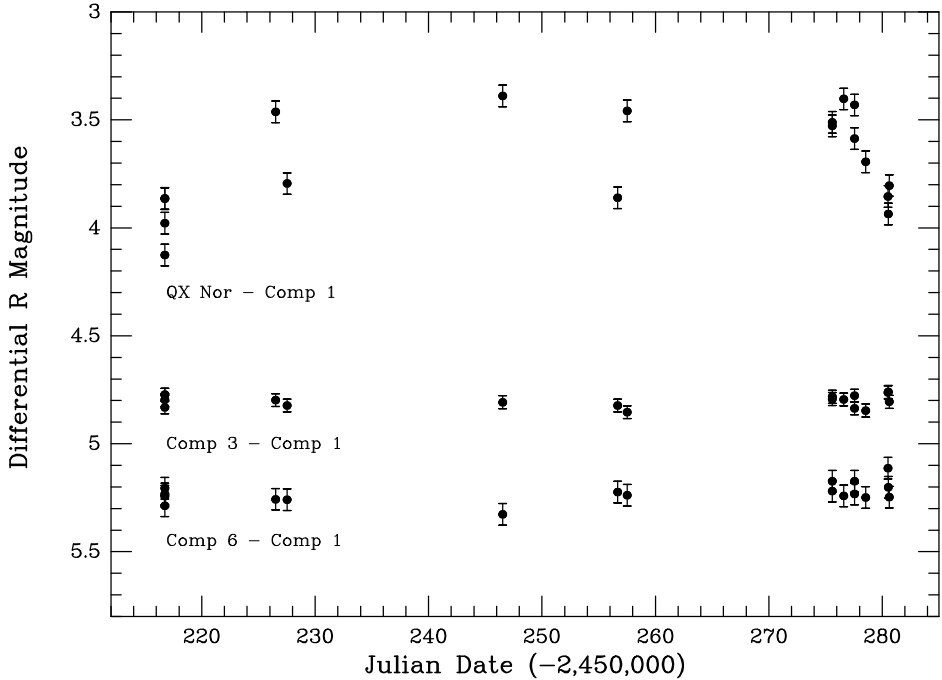}{10.0cm}{-90}{50}{50}{-216}{298}
 \caption[]{Differential $R$ band lightcurve of QX~Nor and
 two comparison stars of similar brightness. The relative magnitudes of the
 comparison stars have been shifted fainter by 1.5~mag to separate
 the lightcurves. The 1$\sigma$ error bars have been derived from the scatter
 in the comparison lightcurves. Note the evident variability of QX~Nor on a
 timescale of one day.
 \label{f-lc}}
 \end{figure}

\subsection{Near-infrared Photometry}

\begin{figure}[ht]
\plotone{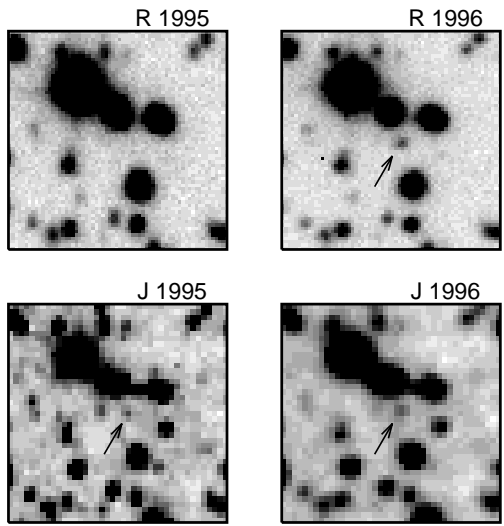}
\caption[]{{\it Top:} Small section of 1995 and
1996 $R$ band images to compare to {\it bottom:} 1995 and 1996 $J$ band
images of the field of QX~Nor. Each $J$ image is the average of nine 15~s
images. QX~Nor ({\it arrow}) is detected in $J$ as a faint source in
{\it both} quiescence and
outburst. North is up and east is to the left.
\label{f-ir}}
\end{figure}

In addition to the optical photometry, near-infrared (IR) 
photometry 
of QX~Nor was performed with the Cerro
Tololo Infrared Imager (CIRIM) on the CTIO 1.5~m telescope
on 1995 June 4 UT ($J$,$Ks$) and 1996 August 28 UT ($J$,$K$). CIRIM uses a 
256$\times$256 HgCdTe NICMOS3 array. All observations were taken in 
the $f/13.5$ mode resulting in a pixel scale
of $0.65\arcsec$~pix$^{-1}$. Four 15~s exposures were coadded to obtain one 
frame. Each observation consisted of a mosaic of nine individual frames, with 
each frame shifted from the previous one by 20$\arcsec$ to 
form a 3$\times$3 grid 
centered on the position of QX~Nor. Dark frames with identical 
integration times and
flat field frames for each filter (derived from observations of an 
illuminated dome
spot) were obtained on each night. First, a mean dark frame was 
subtracted from all 
observations. Next, a sky flat was constructed from a median of the 
scaled object
frames and subtracted from each observation. Finally, the data were 
divided by the 
normalized dome flat. To increase the signal-to-noise and eliminate 
bad pixels, the 
shifted frames were aligned and combined with a bad pixel mask.
A small section of the 1995 and 1996 $J$ frames is displayed
in the bottom panel of Figure~\ref{f-ir}. 

Again, photometry was performed by point spread function fitting with 
DAOPHOT~II. 
The instrumental magnitudes were transformed to the CIT standard system 
through observations of Elias faint infrared standards (Elias
et al. 1982). 
Only $Ks$ (a ``short'' $K$ band filter which reduces the sky background 
contributions) but no $K$
observations were taken in 1995. However, our IR observations of standard
stars in 1996 show that
the instrumental $Ks$ magnitudes are about 0.15 mag fainter than the $K$
magnitudes of the
same objects, so that we can estimate the $K$ magnitude of QX~Nor in 1995
from the $Ks$ observations.
The results for QX~Nor and several local comparison
stars are listed in Table~\ref{t1}. The systematic errors in the IR 
photometry are $\pm 0.1$~mag, and the error in the relative photometry has
been estimated to be between $\pm 0.1$ and $0.2$~mag, again depending on the 
brightness of the stars.  

\section{Results and Discussion}

Figure~\ref{f-fc} shows two 600~s $R$ band exposures obtained in 1995 and 1996. 
Both frames reach a similar limiting magnitude, $R\approx 22$. 
Since QX~Nor was at a mean magnitude of $R=20.2$ during our 1996 post-outburst
observations, 
it has brightened by at least 1.8~mag 
in $R$ from its quiescent state in 1995. 
We also obtained several $I$ band 
observations in 
1996, but have no equivalent data from 1995 for comparison.

Figure~\ref{f-lc} shows the relative $R$ band lightcurve of QX~Nor and two 
local comparison stars. Both comparison lightcurves have been shifted by 
1.5~mag for clarity of display. Comparison~6 is of the same brightness as the
mean level of QX~Nor. Therefore, the 1$\sigma$  
error bars (0.05~mag) derived from the
scatter in the comparison~6 lightcurve have been adopted for QX~Nor.  
The data are insufficient to detect any periodicity, but it is obvious that 
the source is variable from night to night with a full amplitude of
about 0.6~mag. 
  
The bottom panel of Figure~\ref{f-ir} displays a small section of the $J$ 
frames taken in 1995 and 1996.
We detect QX~Nor in {\it both}
quiescence and outburst. A faint source is visible in $J$ but not $R$ in
May 1995. 
Due to its faintness ($J\approx 18$), we took care
to confirm the IR detection of QX~Nor in 1995 and exclude the possibility
of a detector defect. 
The displayed image is a combination of nine shifted frames. The faint object
is seen in each one of those separate frames and similarly in each one of
nine separate $Ks$ frames (not displayed).
The same is true for the 1996 $J$ and $K$ band data. 
The photometry indicates that QX~Nor brightened by about 0.8~mag
in $J$ and $K$ between May 1995 and August 1996. 
Unfortunately, the seeing conditions during the 1996 observations were 
significantly worse than in 1995, which makes a direct comparison of the 
$J$ frames difficult. 
 
Several type~I X-ray bursts (as opposed to transient outbursts) 
have been observed
from X1608-52. Assuming the Eddington luminosity during bursts that show 
photospheric radius expansion (at least one such burst was seen by 
Nakamura et al. 1989) 
puts X1608-52 at a distance of 3.6~kpc.
The extinction to the source can be estimated
from the hydrogen column density ($N_H$) derived from X-ray spectral fits. 
Various $N_H$ values have been suggested for X1608-52 
(Penninx et al. 1989; Mitsuda et al. 
1989; Yoshida et al. 1993) ranging from 
$(1.0 - 2.0)\times 10^{22}$cm$^{-2}$. 
Utilizing the relationship between $A_V$, 
$E(B-V)$, and the column density of Gorenstein (1975) 
leads to $4.5 \leq A_V \leq 9.1$.
A more recent relationship between $A_V$ and $N_H$ based on ROSAT data by
Predehl \& Schmitt (1995) results in $5.6 \leq A_V \leq 11.2$.  
Together with a relationship between extinction and wavelength
(Cardelli, Clayton, \& Mathis  1989), and 
the additional 
constraint of the non-detection in $R$ ($R \gtrsim 22$), we arrive at the
quiescent magnitudes and colors for QX~Nor listed in Table~\ref{t2}.

\begin{figure}[ht]
\plotfiddle{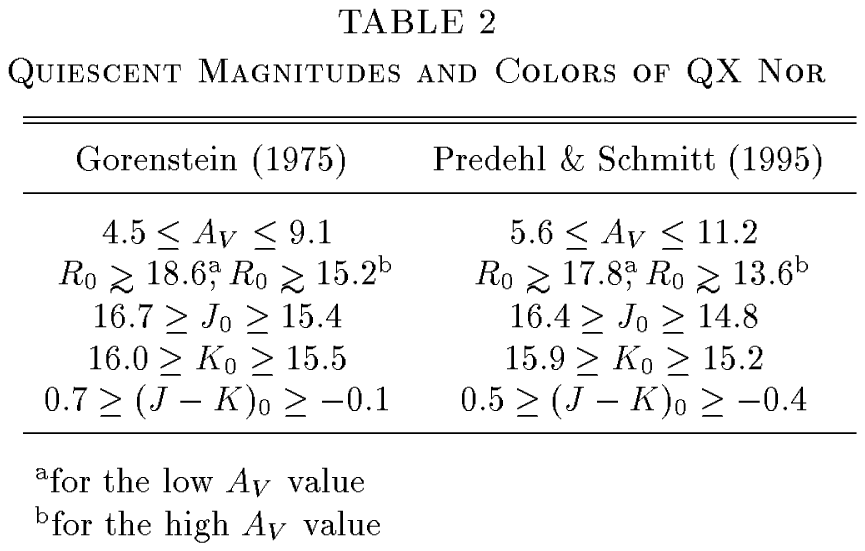}{6.0cm}{0}{100}{100}{-324}{-306}
\end{figure}

For a SXT in quiescence, the main light source is most likely the secondary 
star. There is evidence for a disk in SXTs even in quiescence, but in most 
cases it contributes only 10 - 30\% of the light in the system in the optical
regime (e.g., 
McClintock \& Remillard 1990; Marsh, Robinson, \& Wood 1994).
Additionally, all of our observations are taken
in the red and IR, and may therefore minimize any contribution from a hot
accretion disk. Since we have no 
means of estimating the brightness of any accretion disk in X1608-52, we will 
assume that all light originates from the secondary for the following 
discussion. 
With the exception of GRO J0422+32 (M2V, Filippenko, Matheson, \& Ho 1995) and 
GRO J1655-40 (F5IV, Bailyn et al. 1995),
SXTs have been found to have K dwarf or subgiant companions; 
for example,  Cen~X-4 
(K5 - K7 V/IV, McClintock \& Remillard 1990; Shabaz, Naylor, \& Charles 1993), 
Aql~X-1 (G7 - K3, Thorstensen, Charles, \& Bowyer 1978; K5V, Shabaz et al. 1996),
A0620-00 (K5V, McClintock \& Remillard 1986; K3 - K4 V, Haswell et al. 1993), 
X2023+338 (K0 IV, Casares \& Charles, 1994).
For QX~Nor, the faint magnitudes in quiescence exclude
the possibility of a giant (luminosity class III) companion. 
If the secondary is a main sequence star,
a spectral type of about G0 provides a good match  
with our observed magnitudes and colors for X1608-52 at a 
distance of 3.6~kpc, and implies
$A_V \approx 7 - 8$ for this case. For higher values of $A_V$ 
($A_V \approx 9 - 11$), a spectral type of F0--F5 is still marginally 
consistent with our observations. 
For a K0 type 
companion, X1608-52 has to be at a distance of $\lesssim 3$~kpc, and
for an M0 secondary at  
$d \lesssim 1.5$~kpc. In both of these cases a low $A_V$ from the range of 
possible values is also required. 

\begin{figure}[ht]
\plotfiddle{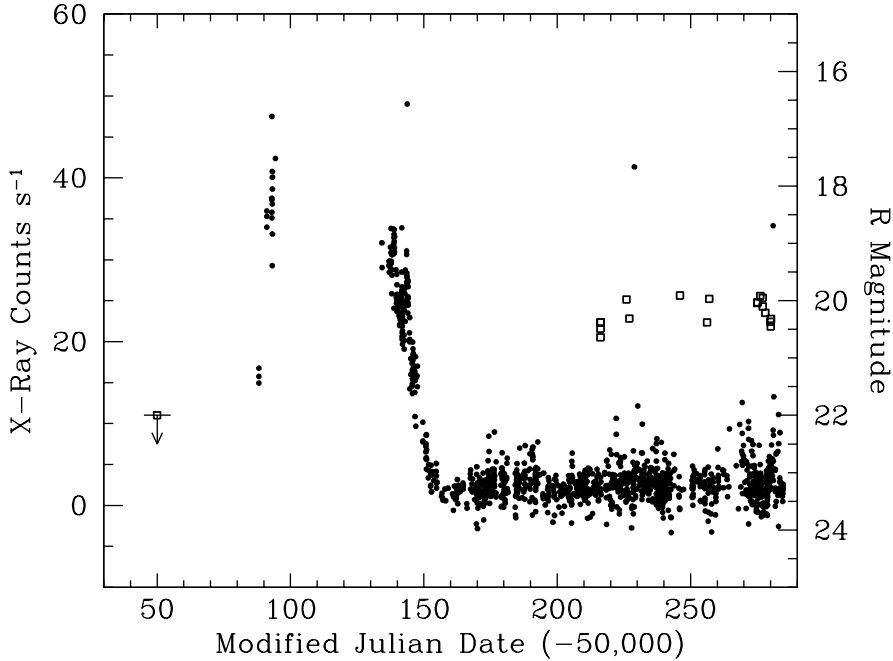}{10.0cm}{-90}{50}{50}{-216}{298}
\caption[]{X-ray lightcurve of the 1996 outburst obtained
by XTE (filled circles).
Also shown are our $R$ band observations (open squares) and the upper limit
to the brightness in $R$ of QX~Nor in quiescence from the 1995 data
(schematically indicated; actual non-detections were on 1995 May 28 UT).
Although the X-rays appear to have reached a quiescent level after the
1996 outburst, the optical data remain well above the quiescent magnitude limit.\label{f-xte}}
\end{figure}

We can compare the two outbursts of X1608-52 for which 
optical and X-ray observations are available. 
Grindlay \& Liller (1978) observed QX~Nor at $I=18.2 \pm 0.2$ about one 
month after the start of the 1977
outburst. Our $I$ band photometry shows QX~Nor at a mean magnitude of 
$I=18.5$ in data taken
5--6 months after the onset of the 1996 outburst.
Figure 2e) in Lochner \& Roussel-Dupre (1994) displays an X-ray lightcurve of
the 1977 outburst taken with the Vela 5B satellite,
which can be compared to the X-ray lightcurve of
the 1996 outburst (obtained from the quicklook results made publicly
available by the ASM/RXTE Team) in our Figure~\ref{f-xte}.

Our optical observations were taken when the X-ray flux of X1608-52 had
returned to an average quiescent level of
$\sim 2.6$~ASM~counts~s$^{-1}$
($\sim 0.03$~Crab).
The Grindlay \& Liller (1978)
observation was obtained {\it during} the outburst when the X-ray flux was
still elevated. According to the X-ray lightcurve in
Lochner \& Roussel-Dupre (1994), the corresponding X-ray flux is
$\sim 30$~Vela~counts~s$^{-1}$ ($\sim 0.75$~Crab). We observe
essentially identical $I$ magnitudes at very different X-ray fluxes,
indicating 
that the X-ray flux is not directly related to the
optical brightness of the system. This is true for most SXTs:
the optical counterparts of SXTs generally  mimic the behavior observed
in X-rays (a steep rise in brightness followed by a gradual decline); 
however, the optical flux falls much more slowly than the X-ray flux,  
so that the optical brightness is still enhanced when the X-ray flux has 
already returned to a quiescent level. 

If we assume a gradual optical flux decay and a similar maximum optical 
brightness in both the 1977 and 1996 outbursts, 
it is surprising that QX~Nor has not 
faded to fainter magnitudes in our 1996 observations, which were obtained 
much later after the onset of an outburst than those 
of Grindlay \& Liller (1978).  
In order to compare the strength of the two outbursts, we tried to estimate 
the maximum X-ray flux reached during both outbursts,
which might be indicative of the maximum optical brightness.
We attempted to fit the 1996 X-ray outburst with an exponential or 
gaussian profile
analogous to the treatment in Lochner \& Roussel-Dupre (1994). 
Unfortunately, there are no data from the ASM at the
time of maximum and we could not achieve a satisfactory fit. An 
exponential profile to the declining branch of the ASM X-ray lightcurve 
requires unreasonably high
maximum count rates. Note that Kaluzienski \& Holt (1977) report that 
the 1977 outburst was characterized by an initial rapid rise followed
by ``a relatively stable plateau level'' that lasted for about a month
and finally a decline phase. It is also evident in Figure e) in Lochner
\& Roussel-Dupre (1994) that an exponential fit does not represent the data  
particularly well.
An outburst profile with a plateau level might also  
provide a better fit to the 1996 outburst. Since we cannot compare the 
two X-ray outbursts, the question of the relation between the two 
$I$ magnitudes must remain unresolved.   
  
Contrary to expectations, we also do not observe any declining trend in 
our 1996 $R$ band data.
Figure~\ref{f-xte} shows our optical observations with
respect to the X-ray outburst.  Although QX~Nor displays some variability,
no obvious overall fading is evident during a time span of two months.
Yet, we know such fading must eventually occur, as these magnitudes are still
significantly brighter than quiescence.
A large number of 
black hole SXTs show secondary and tertiary maxima. One of the best examples
is the lightcurve of V616~Mon (A0620-00), which displays 
the secondary maximum and a plateau-like
state $\sim$200~d after the outburst lasting for about 2~months,
followed by a steep drop in brightness (Whelan et al. 1977). 
It is conceivable that our observations
of QX~Nor were made during such a state. 
If QX~Nor brightens
equally in every band, then our IR observations from 1996 August 28 could 
be an indication
that the system has faded significantly, since it
was only 0.8~mag brighter than in quiescence in $J$ (as opposed to the
earlier 1.8~mag difference in $R$). 

Chen, Livio, \& Gehrels (1993) suggest a complete model for (black hole) 
SXTs based on
the detailed structure in the X-ray and optical decay lightcurves, which  
predicts a specific correlation between X-ray and optical emission.
It would be interesting to investigate whether there are differences in the 
decay lightcurves and the X-ray/optical correlation of systems with neutron
star versus black hole primaries.  
X1608-52 is ideally suited for simultaneous optical/IR and X-ray observations
during a future outburst due to its relatively short outburst recurrence 
time. Only two
outbursts have been observed from Cen~X-4 since 1969, while the outburst 
behavior of Aql~X-1 is peculiar. In addition, QX~Nor could be monitored 
for optical/IR variations in order to obtain its orbital period. 
Both Cen~X-4 and Aql~X-1 have relatively long orbital periods, 15.1~h 
and 19.0~h 
respectively. 
X1608-52 is classified as an Atoll-source and it has been suggested that 
Atoll-sources as a group are characterized by short periods; all other known 
Atoll-source periods are $< 5$~h.
The determination of the orbital period of X1608-52 will show whether
X1608-52 adheres to this general picture and/or whether long orbital 
periods are intimately related to the transient behavior in neutron star 
SXTs.  
 
 \begin{table}
 \dummytable\label{t1}
 \end{table}

 \begin{table}
 \dummytable\label{t2}
 \end{table}

\acknowledgments

We thank Joanne Hughes, Eric Deutsch and Andrew Layden for obtaining 
some of the observations, and     
Don Hoard and Bruce Margon
for reading a draft of this paper and 
providing helpful comments.  This research was in part supported by 
NASA grant NAG5-1630, and  
has made use of the Simbad 
database, operated at CDS, Strasbourg, France.

\end{document}